\documentclass[12pt]{article}
\usepackage{latexsym,amsmath}
\begin{document}
\begin{center}
{\bf \large Dark energy and viscous cosmology}\\

\bigskip

I. Brevik\footnote{Department of Energy and Process Engineering,
Norwegian University of Science and Technology, N-7491 Trondheim,
Norway; email: iver.h.brevik@ntnu.no (corresponding author).} and
O. Gorbunova\footnote{Department of Energy and Process
Engineering, Norwegian University of Science and Technology,
N-7491 Trondheim, Norway. On leave from Tomsk State Pedagogical
University, Tomsk, Russia.}
\end{center}

\begin{abstract}
Singularities in the dark energy universe are discussed, assuming
that there is a bulk viscosity in the cosmic fluid. In particular,
it is shown how the physically natural assumption  of letting the
bulk viscosity be proportional to the scalar expansion in a
spatially flat FRW universe can drive the fluid into the phantom
region ($w<-1$), even if lies in the quintessence region ($w>-1$)
in the non-viscous case.
\end{abstract}

\bigskip

\bigskip


\bigskip


\bigskip
\begin{center}
\today
\end{center}

\section{Introduction}

Recent astrophysical data indicate the presence of a mysterious
kind of energy (ideal fluid with negative pressure), contributing
about 70\% of the total energy of the universe. It is quite
possible that the equation of state parameter $w$ for dark energy
is less than -1. If this is so, the universe shows some very
strange properties such as the future finite singularity called
Big Rip \cite{caldwell03}. In turn, this can lead to the
occurrence of negative entropy \cite{brevik04}. One may expect
that natural effects, most likely of a quantum mechanical origin,
may prevent the Big Rip, as was shown in Ref.~\cite{elizalde04}.
Dependent on which sorts of dark energy are present, various types
of singularities appear; for a classification, see
Ref.~\cite{nojiri05}. In the present letter, we will consider the
role of a {\it bulk viscosity} $\zeta$ in a universe having a Big
Rip singularity. We take the universe to be spatially flat, with
cosmological constant equal to zero. The {\it shear} viscosity
$\eta$ will be put equal to zero, in conformity with usual
practice.  The latter assumption is however a non-trivial point,
since the shear viscosity is usually so much greater than the bulk
viscosity. At least, this is so in the  early universe. For
instance, in the plasma era after the time of recombination ($T
\approx 4000$ K), the ratio $\eta/\zeta$ as calculated from
kinetic theory is as large as about $10^{12}$ \cite{brevik94}.
Therefore, even a slight anisotropy in the cosmic fluid would
easily outweigh the influence from the bulk viscosity. We will
below be concerned with the late universe and not the early one.
However, we ought to bear in mind that the neglect of $\eta$ most
likely rests on the tacit assumption about an extreme fine tuning
of the spatial isotropy of the fluid.

In the next section we present the formalism for the FRW universe
when the equation of state is of a general form, $w=w(\rho)$.
Similarly, we admit a general form for the bulk viscosity,
$\zeta=\zeta(\rho)$.  Thereafter we discuss some special cases. Of
main interest appears to be the case where $\zeta$ is proportional
to the scalar expansion $\theta$, $\zeta=\tau \, \theta$, $\tau$
being a constant. Then, it turns out that the barrier $w=-1$
between the quintessence region ($w>-1$) and the phantom region
($w<-1$) can be {\it crossed}, as a consequence of the bulk
viscosity. To our knowledge, this has not been pointed out before.
The case $\zeta =\tau \,\theta$, implying proportionality of the
bulk viscosity to the divergence of the fluid's velocity vector,
is physical natural, and has been considered earlier in an
astrophysical context. Cf., for instance, the review article of
Gr{\o}n \cite{gron90}.

\section{General formalism}

We assume the standard FRW metric,
\begin{equation}
ds^2=-dt^2+a^2(t)\,d{\bf x}^2, \label{1}
\end{equation}
and put the spatial curvature $k$ as well as the cosmological
constant $\Lambda$ equal to zero. The Hubble parameter is
$H=\dot{a}/a$, and the scalar expansion is $\theta \equiv
{U^\mu}_{;\mu}=3H$, $U^\mu$  being the four-velocity of the fluid.
We assume the equation of state in the form
\begin{equation}
p=w(\rho)\rho, \label{2}
\end{equation}
where the thermodynamical variable $w(\rho)$ is an arbitrary
function of the density. Similarly, we assume that the bulk
viscosity is arbitrary, $\zeta=\zeta(\rho)$. The Friedmann
equations are (cf., for instance, Ref.~\cite{brevik94})
\begin{equation}
\theta^2=24\pi G \rho, \label{3},
\end{equation}
\begin{equation}
\dot {\rho}+(\rho+p)\theta=\zeta\,\theta^2, \label{4}
\end{equation}
\begin{equation}
\frac{\ddot a}{a}+\frac{\theta^2}{18}=-4\pi G (p-\zeta\,\theta).
\label{5}
\end{equation}
The effective pressure is $\tilde{p}=p-\zeta\,\theta$. On
thermodynamical grounds, in conventional physics $\zeta$ has to be
a positive quantity. This is a consequence of the positive sign of
the entropy change in an irreversible process (cf., for instance,
Ref.~\cite{landau87}). We shall assume that $\zeta>0$ also here,
although although strictly speaking the usual thermodynamical
relationships ought to be taken with some care when dealing with
bizarre systems like a phantom fluid. The effect of the bulk
viscosity is accordingly to reduce the thermodynamic pressure in
the fluid. (The viscous Friedmann equations above were considered
in a Cardy-Verlinde entropy context in Ref.~\cite{brevik02}.) Note
that due to the time dependence of the viscosity the Friedmann
equations above may also be interpreted as coming from a
modification of gravity (for a recent discussion and a list of
references, see Refs.~\cite{nojiri03a} and \cite{allemandi04}).

From the above equations we derive the following differential
equation for the scalar expansion:
\begin{equation}
\dot{\theta}+\frac{1}{2}(1+w)\theta^2-12\pi G\,\zeta\,\theta=0
\label{6}
\end{equation}
which, in view of the relationship $\dot{\theta}=\sqrt{6\pi
G}\dot{\rho}/\sqrt{\rho}$, can be written as a differential
equation for $\rho$. Since we will be interested in the region
around $w=-1$, it is convenient, following Nojiri et al.
\cite{nojiri05}, to introduce the function $f(\rho)$ defined by
\begin{equation}
1+w(\rho)=-f(\rho)/\rho, \label{7}
\end{equation}
implying that $f(\rho)=0$ for a "vacuum" fluid. Then, the equation
takes the form
\begin{equation}
\dot{\rho}-\sqrt{24\pi G\rho}\,f(\rho)-24\pi G\zeta(\rho)\rho=0.
\label{8}
\end{equation}
We shall be interested in the development of the late universe,
from $t=t_0$ onwards. For simplicity, we put $t_0=0$. The
corresponding starting value of $\rho$ will be denoted by
$\rho_0$. From Eq.~(\ref{8}) we obtain
\begin{equation}
t=\frac{1}{\sqrt{24\pi G}}\int_{\rho_0}^{\rho}
\frac{d\rho}{\sqrt{\rho}f(\rho)\left[1+\sqrt{24\pi
G}\,\zeta(\rho)\sqrt{\rho}/f(\rho)\right]}. \label{9}
\end{equation}
This is the general relation between the cosmological time $t\;
(>0)$ and the density $\rho$.

\section{The case when $w$ is a constant}

This is thermodynamically the simplest case. Let us put
\begin{equation}
f(\rho)=\alpha \rho, \label{10}
\end{equation}
where $\alpha$ is a constant. This means that $p \equiv
w\rho=-(1+\alpha)\rho$. We investigate in the following some
different choices for $\zeta$.

{\it (i)\;  $\zeta =0$}. In this non-viscous case
 Eq.~(\ref{9}) yields
\begin{equation}
t=\frac{1}{\sqrt{24\pi
G}}\frac{2}{\alpha}\left(\frac{1}{\sqrt{\rho_0}}
-\frac{1}{\sqrt{\rho}}\right)
. \label{11}
\end{equation}
Thus, if $\alpha >0$ a finite value of the time $t$ is compatible
with a Big Rip singularity ($\rho =\infty$). This is the
conventional phantom case, corresponding to $w<-1$. The variation
of the density with time is
\begin{equation}
\rho(t)=\rho_0\left( 1-\frac{1}{2}\alpha \theta_0t\right)^{-2},
\label{12}
\end{equation}
where the initial scalar expansion is
\begin{equation}
\theta_0=\sqrt{24\pi G\rho_0}. \label{13}
\end{equation}
Similarly, the scalar expansion is
\begin{equation}
\theta(t)=\theta_0\left(1-\frac{1}{2}\alpha \theta_0t\right)^{-1},
\label{14}
\end{equation}
and the scale factor is
\begin{equation}
a(t)=a_0\left(1-\frac{1}{2}\alpha \theta_0t\right)^{-2/3\alpha}.
\label{15}
\end{equation}
One may note here that $\rho/\rho_0=(a/a_0)^{3\alpha}$. Thus all
the quantities $\rho(t), \theta(t), a(t)$ diverge at the Big Rip.
The above expressions are in agreement with earlier results
\cite{caldwell03,brevik94}. It ought to be mentioned that for
$w<-1$ the dark energy behaves in close analogy with quantum
fields \cite{nojiri03}.

{\it (ii)\; $\zeta$ equal to a constant}. Equation (\ref{9})
yields in this case
\begin{equation}
t=\frac{1}{\sqrt{24\pi G}}\frac{1}{\alpha}\int_{\rho_0}^\rho
\frac{d\rho}{ \rho^{3/2}\left[1+\sqrt{24\pi G}\,\zeta /(\alpha
\sqrt{\rho})\right]}. \label{16}
\end{equation}
It is however in this case most convenient to go back to
Eq.~(\ref{6}) and solve it with respect to $\theta$ (cf.
Ref.~\cite{brevik94}):
\begin{equation}
\theta(t)=\frac{\theta_0e^{t/t_c}}{1-\frac{1}{2}\alpha
\theta_0t_c(e^{t/t_c}-1)}, \label{17}
\end{equation}
where
\begin{equation}
t_c=(12\pi G\zeta)^{-1}. \label{18}
\end{equation}
Correspondingly,
\begin{equation}
a(t)=a_0\left[ 1-\frac{1}{2}\alpha
\theta_0t_c(e^{t/t_c}-1)\right]^{-2/3\alpha}. \label{19}
\end{equation}
From Eq.~(\ref{3}) then
\begin{equation}
\rho(t)=\rho_0\frac{e^{2t/t_c}}{\left[ 1-\frac{1}{2}\alpha
\theta_0t_c(e^{t/t_c}-1)\right]^2}. \label{20}
\end{equation}
Thus, at the time $t=t_s$, where
\begin{equation}
t_s=t_c\ln \left[1+\frac{2}{\alpha \theta_0t_c}\right], \label{21}
\end{equation}
there occurs a Big Rip; both $\theta(t), a(t)$ and $\rho(t)$
diverge. As before, $\alpha$ has to be a positive quantity,
corresponding to $w=-1-\alpha <-1$. When $\zeta \rightarrow 0$,
the results of the previous case {\it (i)} are recovered.

{\it (iii) The case $\zeta= \tau \theta$.}  This is the most
interesting case. As $\zeta$ is assumed to be positive, the
constant $\tau$ has to be positive. In view of Eq.~(\ref{3}) we
can alternatively write the bulk viscosity as a function of the
density:
\begin{equation}
\zeta(\rho)=\tau \sqrt{24\pi G \rho}. \label{22}
\end{equation}
Equation (\ref{9}) yields in this case
\begin{equation}
t=\frac{1}{\sqrt{24\pi G}}\,\frac{2}{\alpha +24\pi G
\tau}\left(\frac{1}{\sqrt{\rho_0}}-\frac{1}{\sqrt{\rho}}\right).
\label{23}
\end{equation}
From this condition it follows that the condition for a Big Rip
($\rho=\infty$) to occur in a finite time $t$ is that the
prefactor is positive,
\begin{equation}
\alpha+24\pi G\tau >0. \label{24}
\end{equation}
This is the most important result of the present paper. Even if we
start from a situation where $w>-1$ (i.e., $\alpha <0$),
corresponding to the quintessence region for an ideal fluid, the
presence of a sufficiently large bulk viscosity will make the
condition (\ref{24}) satisfied and thus drive the fluid into the
Big Rip singularity. Somewhat surprisingly, it is physically the
reduction of the thermodynamical pressure generated by the
viscosity which in turn causes the barrier $w=-1$ to be crossed.

The system is actually most easily analyzed by going back to
Eq.~(\ref{6}) for the scalar expansion. Introducing the effective
new parameter
\begin{equation}
\tilde{\alpha}=\alpha+24\pi G\tau, \label{25}
\end{equation}
we can write the equation as
\begin{equation}
\dot{\theta}-\frac{1}{2}\tilde{\alpha}\,\theta^2=0. \label{26}
\end{equation}
From this it follows that $\rho(t), \theta(t)$ and $a(t)$ can
actually be found from the non-viscous expressions
(\ref{12})-(\ref{15}), only with the replacement $\alpha
\rightarrow \tilde{\alpha}$.

In general, the existence of viscosity coefficients in a fluid is
due to the thermodynamic irreversibility of the motion. If the
deviation from reversibility is small, the momentum transfer
between various parts of the fluid (i.e., the stress tensor) can
be taken to be linearly dependent on the velocity derivatives
$\partial_k u_i$. This is the conventional case, corresponding to
constant viscosity coefficients. The present case $\zeta = \tau
\theta$ means that we go step further; it means that the momentum
transfer involves second order quantities in the deviation from
reversibility, still maintaining the scalar property of $\zeta$.
As mentioned above, the ansatz $\zeta =\tau \theta$ has been
considered earlier in a cosmological context \cite{gron90}.

\section{Remarks on the case $w=w(\rho)$}

In this general case we have to go back to the expression
(\ref{9}). As the ansatz $\zeta=\tau \theta$ appears to be that of
main interest, we consider this case first. From Eq.~(\ref{9}) we
get
\begin{equation}
t=\frac{1}{\sqrt{24\pi G}}\int_{\rho_0}^\rho
\frac{d\rho}{\sqrt{\rho}f(\rho)[1+24\pi G\tau \rho /f(\rho)]}.
\label{27}
\end{equation}
Thus, if $f(\rho) \rightarrow \alpha \rho$ in the limit of large
$\rho$, the Big Rip singularity is allowed. The necessary
condition on the values of $\alpha$ and $\tau$ are again given by
Eq.~(\ref{24}).

Another choice for $f(\rho)$ that may seem natural, is to take
\begin{equation}
f(\rho)=A\rho^\beta \label{28}
\end{equation}
for all $\rho$, where $A$ and $\beta$ are constants (cf.
Refs.~\cite{nojiri05,nojiri04,stefancic04}). Thus
$p=-\rho-A\rho^\beta$. We shall here take $A$ and $\beta$ to be
positive. From Eq.~(\ref{27}) we get
\begin{equation}
t=\frac{1}{\sqrt{24\pi
G}}\frac{1}{A}\int_{\rho_0}^\rho\frac{\rho^{-\beta-1/2}}{1+(24\pi
G\tau/A)\rho^{1-\beta}}. \label{29}
\end{equation}
If $\beta >1$ the influence from viscosity is seen to fade away
for large $\rho$, and the Big Rip is allowed. If $\beta <1$ the
viscosity term becomes dominant for large $\rho$, and if we
perform the integration over a region $\rho \in [\rho_0,\rho]$
where $\rho$ is large, we get
\begin{equation}
t \approx \frac{2}{(24\pi
G)^{3/2}}\frac{1}{\tau}\left(\frac{1}{\sqrt{\rho_0}}-\frac{1}{\sqrt{\rho}}\right).\label{30}
\end{equation}
We can thus have Big Rip also when $\beta <1$. Close to the
singularity, the time $t$ depends on viscosity through the factor
$\tau$ but is independent  of $A$ and $\beta$. Recall that the
formalism so far in this section assumes that $\zeta =\tau
\theta$.

The final point that we shall deal with, is the following
question: what form for the function $f(\rho)$ does the ansatz
\begin{equation}
a(t)=a_0\left( \frac{t}{t_s-t}\right)^n \label{31}
\end{equation}
correspond to? Here $n$ is a positive constant and $t_s$ is the
Big Rip time. The ansatz (\ref{31}) is simple and suggestive, and
has been dealt with before in the non-viscous case \cite{nojiri05}
. From Eqs.~(\ref{31}) and (\ref{3}) we get
\begin{equation}
\theta(t)=3n\left(\frac{1}{t}+\frac{1}{t_s-t}\right), \label{32}
\end{equation}
\begin{equation}
\rho(t)=\frac{3n^2}{8\pi
G}\left(\frac{1}{t}+\frac{1}{t_s-t}\right)^2, \label{33}
\end{equation}
leading to
\begin{equation}
\dot{\rho}(t)=\pm 2\rho \left\{ \frac{8\pi
G}{3}\frac{\rho}{n^2}-
\frac{4}{nt_s}\sqrt{\frac{8\pi G}{3}\rho}
\right\}^{1/2}. \label{34}
\end{equation}The doubly-valued equation of state is typical for a
first-order transition (cf. a more extensive discussion on this
point in Ref.~\cite{nojiri05}). These equations are the same as in
the non-viscous case. But there is a difference arising from
Eq.~(\ref{4}): we get
\begin{equation}
f(\rho)=\pm \frac{2\rho}{3n}\left\{
1-\frac{4n}{t_s}\sqrt{\frac{3}{8\pi G\rho}}\right\}^{1/2}-3\zeta
(\rho)\sqrt{\frac{8\pi G}{3}\rho}, \label{35}
\end{equation}
showing the influence from viscosity in the last term. Note that
$\zeta=\zeta(\rho)$ is here an arbitrary function. Thus, the
initial simple ansatz (\ref{31}) for $a(t)$ leads to a rather
complicated form for $f(\rho)$.

It would be interesting to generalize the present work taking into
account quantum effects together with viscosity. This may be done
in close analogy with previous attempts in this direction, for an
asymptotically de Sitter universe \cite{brevik99}.

\bigskip

\noindent {\bf Acknowledgment}  We thank Sergei D. Odintsov for
valuable discussions and suggestions.

\end{document}